\documentclass[twocolumn]{aastex701}
\usepackage{apjfonts}
\usepackage{rotating}
\usepackage{amsmath, color,graphicx,tablefootnote,subfigure}
\usepackage{graphics,subfigure,hyperref,float}
\usepackage[rightcaption]{sidecap}

\newcommand{\D}{$^{\dagger}$}

\newcommand{\W}{$\lambda$}

\newcommand{\HII}{H$\,${\sc ii}}
\def\hii  {\ifmmode{{\rm H}{\rm \small II}}\else{H\ts {\scriptsize II}}\fi}

\def\pfrac#1#2{\left( \frac{#1}{#2} \right)}
\def\iso#1#2{\mbox{${}^{#2}{\rm #1}$}}
\def\he#1{\iso{He}{#1}}
\def\li#1{\iso{Li}{#1}}
\def\be#1{\iso{Be}{#1}}

\def\Yp{Y_{\rm p}}

\def\omb{\Omega_{\rm B} h^2}

\def\beq{\begin{equation}}
\def\eeq{\end{equation}}
\def\beqar{\begin{eqnarray}}
\def\eeqar{\end{eqnarray}}

\shorttitle{The LBT $Y_{\rm p}$ Project V}
\shortauthors{Yeh et al.}

\begin{document}

\title{The LBT $\Yp$ Project V: Cosmological Implications of a New Determination of Primordial \he4}

%
%
\author[0000-0002-4137-5306]{Tsung-Han Yeh}
\affiliation{TRIUMF, 4004 Wesbrook Mall, Vancouver, BC V6T 2A3, Canada}
\email{thyeh@triumf.ca}
\author[0000-0001-7201-5998]{Keith A. Olive}
\affiliation{William I. Fine Theoretical Physics Institute, School of Physics and Astronomy, University of Minnesota, Minneapolis, MN 55455, USA}
\email{olive@umn.edu}
\author[0000-0002-4188-7141]{Brian D. Fields}
\affiliation{Department of Astronomy, University of Illinois, Urbana, IL 61801}
\affiliation{Department of Physics, University of Illinois, Urbana IL 61801}
\affiliation{Illinois Center for Advanced Studies of the Universe}
\email{bdfields@illinois.edu}
\author[0009-0006-2077-2552]{Erik Aver}
\affiliation{Department of Physics, Gonzaga University, 502 E Boone Ave., Spokane, WA, 99258}
\email{aver@gonzaga.edu}
\author[0000-0003-1435-3053]{Richard W.\ Pogge}
\affiliation{Department of Astronomy, The Ohio State University, 140 W 18th Ave., Columbus, OH, 43210}
\affiliation{Center for Cosmology \& AstroParticle Physics, The Ohio State University, 191 West Woodruff Avenue, Columbus, OH 43210}
\email{pogge.1@osu.edu}
\author[0000-0002-0361-8223]{Noah S.\ J.\ Rogers}
\affiliation{Center for Interdisciplinary Exploration and Research in Astrophysics (CIERA), Northwestern University, 1800 Sherman Avenue, Evanston, IL 60201, USA}
\email{noah.rogers@northwestern.edu}
\author[0000-0003-0605-8732]{Evan D.\ Skillman}
\affiliation{Minnesota Institute for Astrophysics, University of Minnesota, 116 Church St. SE, Minneapolis, MN 55455}
\email{skill001@umn.edu}
\author[0000-0003-4912-5157]{Miqaela K.\ Weller}
\affiliation{Department of Astronomy, The Ohio State University, 140 W 18th Ave., Columbus, OH, 43210}
\email{weller.133@buckeyemail.osu.edu}
%


\begin{abstract}
The primordial abundance of \he4 plays a central role in big-bang nucleosynthesis (BBN) and in the cosmic microwave background (CMB). The LBT $\Yp$ Project's new measurement of the primordial \he4 mass fraction $\Yp =0.2458 \pm 0.0013$ is the most precise determination to date.  In this paper, we combine our new $\Yp$ value with the latest primordial deuterium measurement, and assess the consequences for cosmology.
For Standard BBN, where the number of light neutrino species is fixed at $N_\nu=3$, the single free parameter is the cosmic baryon density; the CMB measures this independently, with results consistent with each other.  Combining $\Yp$, D/H, BBN, and the CMB, gives the cosmic baryon-to-photon ratio $\eta = (6.120 \pm 0.038) \times 10^{-10}$, corresponding to a baryon density parameter $\omb = 0.02236 \pm 0.00014$.  
We then allow $N_\nu$ to vary and thus measure relativistic species present during nucleosynthesis.    We find $\eta = (6.101 \pm 0.044) \times 10^{-10}$ or
$\omb = 0.02229 \pm 0. 00016$, and $N_\nu = 2.925 \pm 0.082$, and for $N_\nu \ge 3$, 
$\Delta N_\nu= N_\nu-3 \le 0.125$ (95\% CL) during BBN and the CMB.  
Our results demonstrate consistency with the Standard Model of particle physics, and with the standard cosmology that links BBN at $\sim 1 \ \rm sec$ and the CMB at $\sim 400,000$ yr.
\end{abstract}



\section{Introduction}
\label{sec:intro}

There are relatively few direct probes of the very early Universe. Observations of the cosmic microwave background (CMB), and in particular the power spectrum of anisotropies, provides precision information corresponding to a redshift of $z\simeq 1100$. Big bang nucleosynthesis (BBN) and observational determinations of the abundances of deuterium and $\he4$ provide information on the state of the Universe at a redshift of $z\sim 10^{10}$. If the origin of the CMB anisotropies is associated with a period of inflation in the early universe, then these observations also provide information on the very earliest moments in the evolution of the Universe at a redshift of $z \sim 10^{28}$.
However, BBN provides the earliest probe where {\em known} physics can be tested \citep[see e.g.,][]{Sarkar:1995dd,Cyburt2005,Jedamzik:2009uy,Pospelov:2010hj,Yeh:2022heq}. BBN is operative at a temperature of roughly 1 MeV, where the theories of weak and nuclear interactions are well tested. As a result, any physics beyond the Standard Model can be constrained provided we obtain accurate abundance measurements. 

There are four light element isotopes produced in BBN: D, \he3, \he4 and \li7. To test BBN, we must have accurate abundance measurements that can be associated with a primordial abundance.  At present, the predicted abundances of two of these isotopes lend themselves to observational tests: D and \he4. Generally this requires an observation providing either emission or absorption line data from very metal-poor environments where the effects of stellar contamination are negligible.  

Deuterium is observed in high redshift ($z\sim 3$) quasar absorption systems. It is destroyed as it passes through stars in the process of chemical evolution, and in simple models one can show that the deuterium abundance is exponentially sensitive to metallicity \citep{1975ApJ...201L..51O} with ${\rm D} \propto {\rm D_p} e^{-Z/Z_*}$, where D$_{\rm p}$
is the primordial abundance, Z is the metallicity, and $Z_* \sim Z_\odot$. At low metallicity ($Z \sim 10^{-2} Z_\odot$), we expect ${\rm D} \simeq {\rm D_p}$.
While there are only a handful of useful systems, the accuracy of the measurements is exceptional. Based on 12 objects \citep{cooke2014, cooke2016, cooke2018, riemer2015, riemer2017, balashev2016,Zavarygin:2018,guarneri2024MNRAS.529..839G,kislitsyn2024MNRAS.528.4068K}, 
\beq
\left(\frac{\rm D}{\rm H}\right)_{\rm p} = (2.513 \pm 0.028) \times 10^{-5} \, . 
\label{DHnew}
\eeq
As we discuss in more detail below, the theoretical accuracy due to uncertainties in nuclear cross sections involving deuterium lags the current observational uncertainty. Notably, the $d(d,p)t$ and $d(d,n)\he3$ reactions
have large uncertainties preventing
a more accurate prediction \citep{2021Yeh}. Nevertheless, the standard BBN (hereafter SBBN) prediction for D/H, 
assuming the baryon density determined from CMB observations \citep{Planck:2018vyg} is in excellent agreement with the observational determination of D/H. 

In contrast, \he4 is observed in relatively low redshift \HII\ regions, though at low metallicity. Up until now, there have been very few observations of objects with sufficiently low metallicity, such that the effects of stellar contamination could be ignored. In contrast to D/H which decreases with stellar processing, the helium abundance $y= n_{\rm He}/n_{\rm H}$, or mass fraction $Y = \rho_{\rm He}/\rho_{\rm B}$, increases with time and stellar processing. In this case, assuming a linear relation between $Y$ and metallicity with O/H as a surrogate for metallicity $Z$, a linear regression can be calculated allowing one to extrapolate to zero metallicity and thereby obtain the {\em primordial} helium abundance, $Y_{\rm p}$ \citep{peimbert1974}. 
In addition to the lack of sufficiently low-metallicity targets, the determination of the helium abundance from a limited set of emission line fluxes is subject to substantial systematic uncertainties \citep{olive2001,olive2004}. Some of these systematics are related to degeneracies among the underlying physical characteristics of the \HII\ region, such as electron density and temperature. Observations with additional He and H emission line data can be useful for breaking these degeneracies \citep{olive2001, olive2004, Luridiana:2003jy, Peimbert:2005wt, Izotov:2007ed, aver2010,aver2011,aver2012,aver2015} leading to an improved and realistic determination of the uncertainty in the helium abundance. For further discussion, see Paper I of this series \citep{Skillman2025}. 

Even with precision determinations of the helium abundance in individual \HII\ regions, we still have a residual systematic uncertainty stemming from stellar nucleosynthesis that contributes to the total helium abundance. A linear regression assumes that $y$ tracks O/H linearly and that uncertainty is typically not accounted for. Thus, one of the goals of the LBT $Y_{\rm p}$ project \citep{Skillman2025,Rogers2025,Weller2025,Aver2025} is to obtain a sufficiently large sample of extremely metal-poor \HII\ regions
so as not to be sensitive to or rely on a regression to determine $Y_{\rm p}$. The results of this project have enabled us to obtain helium abundances from 15 \HII\ regions with O/H $\le 4 \times 10^{-5}$. 
To put this in perspective, in \cite{aver2015}, the value of $Y_P = 0.2449 \pm 0.0040$ based on a regression was established using 15 objects with a metallicity ${\rm O/H} \lesssim 15 \times 10^{-5}$. Through several improvements in the analysis \citep{Aver:2020fon,Aver:2021rwi} and the addition of two low-metallicity objects, this was improved to $Y_P = 0.2448 \pm 0.0033$ \citep{Aver:2021rwi}. The LBT $Y_{\rm p}$ project resulted in observations of 41 new observations with O/H $< 14.5 \times 10^{-5}$ of which 15 objects have O/H $\le 4 \times 10^{-4}$. Concentrating on the these very low-metallicity objects (which show no sign of a correlation with O/H),
the primordial helium abundance determined from their mean value is \citep{Aver2025}
\beq
Y_{\rm p} = 0.2458 \pm 0.0013 \, .
\label{Ypnew}
\eeq
This corresponds to a helium-to-hydrogen ratio of
\beq
\pfrac{\he4}{\rm H}_{\rm p} = y_{\rm p} = 0.08146 \pm 0.00058 \, ,
\eeq
which is actually the quantity coming directly from the observations\footnote{Strictly speaking, spectroscopic measurements determine the {\em elemental} He/H ratio, but interstellar measurements show $\rm \he3/H \sim 10^{-5}$ \citep{Bania2002}, so that in practice $y_{\rm p}$ measures \he4/H.}.
The two are connected by $Y_{\rm p}= 4y_{\rm p}/(1+4y_{\rm p})$ because strictly speaking, $Y_{\rm p}$ here is the baryon fraction of \he4.
Further details of this result will be discussed in Section \ref{sec:new}.  

This result has implications beyond the concordance between BBN theory and observations (from the CMB and the light elements). It is well known that the helium abundance has the ability to place strong constraints on physics beyond the Standard Model. Indeed, the helium abundance is sensitive to the expansion rate of the Universe at the time of BBN which in turn depends on the number of relativistic degrees of freedom \citep{Hoyle1964,Shvartsman1969,Peebles1971,SSG1977}. The latter can often be expressed in terms of the number of neutrino flavors, $N_\nu$. In the Standard Model, $N_\nu = 3$, and the number of neutrinos contributing to the width of the Z gauge boson was measured at LEP to be $N_\nu = 2.9963 \pm 0.0074$ \citep{Janot:2019oyi}.
Thus, in SBBN, we follow the Standard Model to fix $N_\nu = 3$;
note that this corresponds to an effective number of neutrinos, $N_{\rm eff} = 3.044$ \citep{{Drewes:2024wbw}}, when accounting for the residual $e^+ e^-$ annihilations before neutrino decoupling. When $N_\nu$ is not held fixed, which we refer to as NBBN, minimization of the combined BBN/CMB/Observational likelihood function can determine a best fit value for $N_\nu$ \citep{Yeh:2022heq}.
The previous result based on the helium abundance in \cite{Aver:2021rwi} was $N_\nu = 2.898 \pm 0.141$. In what follows, we will see that the best fit value of $N_\nu$ moves closer to the Standard Model value with a significantly smaller uncertainty. 

The remaining two element isotopes, \he3 and \li7, are not currently useful for BBN analysis. \he3 is both produced and destroyed in stars \citep{Olive:1994fq,Dearborn:1995ex,Olive:1996tt}, with low-mass stars being net producers of \he3 and high-mass stars being net destroyers of \he3. Without knowing the true star formation history as well as the stellar mass distribution,  extracting a primordial abundance from available observations is both difficult and model-dependent \citep{Vangioni-Flam:2002cvh}. 
Similarly, stars may have a significant effect on the abundance of \li7. Low-mass halo dwarf stars show a uniform abundance of \li7 over a wide range of stellar parameters \citep{Spite:1982dd}. In particular, for iron abundances [Fe/H] $< -1.5$ (that is roughly 1/30 the solar value), the \li7 abundance appears independent of metallicity.  This is often associated with the primordial abundance \citep{Yang:1983gn}. The lack of dispersion in the \li7 data, coupled with supposed observations of \li6 \citep{Smith:1993zzg}, strengthened the connection between these observations and BBN.  However, the BBN prediction is significantly higher than the lithium plateau \citep{Cyburt:2008kw}. 
Furthermore, more recent observations, particularly at low metallicity, show signs of dispersion \citep{Sbordone:2010zi,Bonifacio:2018hrc,Aguado:2019egq} (expected if Li is depleted in stars), and the \li6 observations have been called into question \citep{Wang:2021urb}. 
It is now unclear the degree to which \li7 has been destroyed in its host star \citep{Fields:2022mpw} making it unsuitable for BBN analysis.

While the improved accuracy in $Y_{\rm p}$ in Eq.~(\ref{Ypnew}) now exceeds the relative accuracy in D/H given in Eq.~(\ref{DHnew}), as we discuss in more detail below, to fully test SBBN, we need to aim for further improvement in the uncertainty in $Y_{\rm p}$. The goal of adding more observations of very low-metallicity objects now appears attainable. 
In what follows, we provide a brief summary of our SBBN analysis in Section \ref{sec:SBBN}. This analysis combines BBN likelihood functions convolved with Planck CMB likelihoods and with the observational likelihoods based on Eqs.~(\ref{DHnew}) and (\ref{Ypnew}). In Section \ref{sec:new}, we summarize the new LBT observations that lead to Eq.~(\ref{Ypnew}). Our results are given in Section \ref{sec:results} where we discuss not only the concordance of SBBN,
but also the constraints on physics beyond the Standard Model as parameterized as a limit on $N_\nu$. Our conclusions are summarized in Section \ref{sec:sum}.

\section{BBN Calculations}
\label{sec:SBBN}

Our BBN calculations are based on the code used in our previous work \citep{Fields:2019pfx,2021Yeh,Yeh:2022heq}.
For further details, see \citet{Cyburt:2001pp,Cyburt:2015mya,Fields:2019pfx}.
For results from other BBN codes, see e.g.~\citet{Iocco2009,Pitrou2018}.
Here we summarize the key ingredients and highlight the small differences from our earlier work.
For SBBN, the one free parameter is the baryon density, measured by the baryon-to-photon ratio
\beq
\eta = \frac{n_{\rm b}}{n_\gamma} \equiv 10^{-10} \ \eta_{10}  \ \ ,
\eeq
which is the ratio of the baryon and photon number densities.
In the standard cosmology, $\eta$ does not change from the end of BBN, to the CMB recombination epoch, to today.
Equivalently, the cosmic baryon content can be expressed in terms of the baryon density parameter; the two are related by
\begin{eqnarray}
    \omb & = & 3.6529 \times 10^{-3} \pfrac{T_0}{2.7255 \ \rm K}^3 \eta_{10} \nonumber \\
    & &  \times \left[ 1-7.131\times 10^{-3} (Y_p-0.245) \right] \ \ .
    \label{ombeta}
\end{eqnarray}
The second line in this expression gives the weak dependence of the baryon density on the primordial \he4 abundance \citep{Steigman2006,Fields:2019pfx}.  This reflects small differences in the mass per baryon for hydrogen and \he4, arising from the \he4 nuclear binding energy.

The initial conditions for SBBN are set when the temperature of the Universe is of order 1 MeV. At that time, the Universe is dominated by radiation
with an energy density given by
\beq
\rho_{\rm rad} = \rho_\gamma + \rho_e + \rho_\nu =  \frac{\pi^2}{30} \left(2 + \frac72 + \frac74 N_\nu \right) T^4 \, ,
\eeq
where $N_\nu$ is the total number of neutrino flavors (relativistic at temperature $T$).
Standard BBN adopts $N_\nu = 3$ from the Standard Model of particle physics.
Any additional relativistic particle species present during BBN would contribute to $\rho_{\rm rad}$ and can be expressed in terms of the equivalent number $N_\nu$.

In our calculations, the thermonuclear reaction rates are the same as in \citet{Yeh:2022heq}, 
with the exception of the neutron lifetime.
For this, we adopt the most recent Particle Data Group world average
\footnote{https://pdg.lbl.gov/}
\beq
\tau_n = 878.3 \pm 0.4~{\rm s} \ \ .
\label{newtaun}
\eeq
This updates the earlier result $878.4 \pm 0.5$ sec
\citep{ParticleDataGroup:2024cfk}: the central value has barely changed, though the uncertainty is somewhat lower.
We will see that the impact on the predicted \he4 abundance is small, because other reaction uncertainties dominate.

It is useful to see the scalings of the light element abundances to the BBN parameters and reaction rates.  These are:
\begin{widetext}

\beqar
Y_p & = & 0.24672\!\left(\frac{\eta_{10}}{6.120}\right)^{0.039}\!\!\left(\frac{N_\nu}{3.0}\right)^{0.163}\!\!\left(\frac{G_N}{G_{N,0}}\right)^{0.363}\!\!\left(\frac{\tau_n}{878.3  {\rm \ s}}\right)^{0.733}
\label{yfit} 
 \left[ p(n,\gamma)d\right]^{0.005}\left[ d(d,n)\he3\right]^{0.006}\left[ d(d,p)t\right]^{0.005}   \\
\frac{\rm D}{\rm H} &=& 2.493\!\times\! 10^{-5}\!\left(\frac{\eta_{10}}{6.120}\right)^{-1.634}\!\!\left(\frac{N_\nu}{3.0}\right)^{0.405}\!\!\left(\frac{G_N}{G_{N,0}}\right)^{0.989}\!\!\left(\frac{\tau_n}{878.3 {\rm \ s}}\right)^{0.419} \\
& & \times \ \left[ p(n,\gamma)d\right]^{-0.197}\left[ d(d,n)\he3\right]^{-0.535}\left[ d(d,p)t\right]^{-0.462} 
\left[d(p,\gamma)\he3\right]^{-0.344}\left[\he3(n,p)t\right]^{0.024}\left[\he3(d,p)\he4\right]^{-0.015}  \nonumber \\
\frac{\he3}{\rm H} &=& 10.41\!\times\! 10^{-6}\!\left(\frac{\eta_{10}}{6.120}\right)^{-0.571}\!\!\left(\frac{N_\nu}{3.0}\right)^{0.136}\!\!\left(\frac{G_N}{G_{N,0}}\right)^{0.333}\!\!\left(\frac{\tau_n}{878.3 {\rm \ s}}\right)^{0.142}  \\
&& \times \ \left[ p(n,\gamma)d\right]^{0.085}\left[ d(d,n)\he3\right]^{0.196}\left[ d(d,p)t\right]^{-0.256}
\left[d(p,\gamma)\he3\right]^{0.390}\left[\he3(n,p)t\right]^{-0.166}\left[\he3(d,p)\he4\right]^{-0.759}\left[t(d,n)\he4\right]^{-0.008}  \nonumber \\
\frac{\li7}{\rm H} &=& 4.945\!\times\! 10^{-10}\!\left(\frac{\eta_{10}}{6.120}\right)^{2.116}\!\!\left(\frac{N_\nu}{3.0}\right)^{-0.285}\!\!\left(\frac{G_N}{G_{N,0}}\right)^{-0.743}\!\!\left(\frac{\tau_n}{878.3 {\rm \ s}}\right)^{0.432}  \\
&& \times \ \left[ p(n,\gamma)d\right]^{1.309}\left[ d(d,n)\he3\right]^{0.677}\left[ d(d,p)t\right]^{0.062}
\left[d(p, \gamma)\he3\right]^{0.620} \left[\he3(n,p)t\right]^{-0.266} \left[\he3(d,p)\he4\right]^{-0.750} \left[t(d,n)\he4\right]^{-0.020} \nonumber  \\
&& \times \ \left[\he3(\alpha,\gamma)\be7\right]^{0.966}\left[\be7(n,p)\li7\right]^{-0.692}\left[\li7(p,\alpha)\he4\right]^{-0.051}\left[t(\alpha,\gamma)\li7\right]^{0.027}%
\left[\be7(n,\alpha)\he4\right]^{-0.001}\left[\be7(d,p)\he4\he4\right]^{-0.008}  \nonumber
\label{li7fit}
\eeqar
\end{widetext}
in an obvious notation.  

We use a likelihood analysis to compare these BBN theory predictions with observations from both the light elements and the CMB.  This process is explained in detail in \citet{Cyburt:2015mya} and \citet{Fields:2019pfx};
here we summarize the key aspects.
The uncertainties in our BBN code arise from uncertainties in the nuclear reaction cross sections,
and thus the thermonuclear reaction rates.  These are quantified via Monte Carlo calculations, which generate BBN likelihood
function ${\mathcal L}_{\rm BBN}(\eta; X)$ for SBBN with $X \in$ ($\Yp$ or D/H).  When $N_\nu$ is allowed to vary, we have
${\mathcal L}_{\rm NBBN}(\eta, N_\nu; X)$.
For observations of $\Yp$ and deuterium, we use a Gaussian for $\mathcal{L}_{\rm obs}(X)$.
In the results we show, all likelihood functions are normalized so that their peak takes the common
value of 1.

Independent of BBN, the CMB is strongly sensitive to the baryon density, but also to $N_\nu$ and $\Yp$.  The neutrino number $N_{\rm eff}$ affects matter-radiation equality, while the helium abundance affects the number of electrons per baryon.
We create the CMB likelihood from Hermite polynomial fits to the {\em Planck} MCMC chains, for the case of {\small \verb+base_yhe_plikHM_TTTEEE_lowl_lowE_post_lensing+},
which does not impose the BBN relationship between $\Yp$ and the baryon density.
This gives $\mathcal{L}_{\rm CMB}(\eta,\Yp)$ when we fix $N_{\rm eff}$ to the Standard Model value. When $N_{\rm eff}$ is allowed to vary, we use 
the {\em Planck} {\small \verb+base_nnu_yhe_plikHM_TTTEEE_lowl_lowE_post_lensing+} chains, 
to establish $\mathcal{L}_{\rm NCMB}(\eta,N_\nu;\Yp)$.
For each of these cases, we can marginalize over $\Yp$
to find a CMB likelihood for the baryon density.  For example, 
\beq
{\mathcal L}_{\rm CMB}(\eta) \propto \int 
  {\mathcal L}_{\rm CMB}(\eta,Y_{\rm p}) \ dY_{\rm p} \, ,
  \label{LCMB}
\eeq
and a similar result for $\mathcal{L}_{\rm NCMB}(\eta,N_\nu)$.
Alternatively, we can marginalize over baryon density,
\beq
{\mathcal L}_{\rm CMB}(Y_p) \propto \int 
  {\mathcal L}_{\rm CMB}(\eta,Y_p) \ d\eta 
  \label{CMByp}
\eeq
 to obtain the CMB likelihood function  for $\Yp$.

To link theory and observation we combine the likelihoods in a convolution.
For SBBN theory and the light-element observations $X_i \in (\Yp,{\rm D/H})$, 
we have
\beq
{\mathcal L}_{\rm BBN-obs}(\eta) \propto \int 
  {\mathcal L}_{\rm BBN}(\eta;X_i) \
  {\mathcal L}_{\rm obs}(X_i) \ dX_i \, ,
  \label{OBS-eta}
\eeq
which gives the BBN-only distribution in $\eta$.
Turning to the CMB, one approach is to combine BBN theory and CMB observations,
to yield predictions for the light-element abundances:
\beq
{\mathcal L}_{\rm CMB-BBN}(X_i) \propto \int 
  {\mathcal L}_{\rm CMB}(\eta,Y_p) \
  {\mathcal L}_{\rm BBN}(\eta;X_i) \ d\eta \, ,
\label{CMB-BBN}
\eeq
which we then can compare to the observations.
Another approach is to combine BBN and the CMB, including the CMB $\Yp$ dependence,
to obtain a prediction for $\eta$ independent of the light-element abundances:
\beq
{\mathcal L}_{\rm CMB-BBN}(\eta) \propto \int 
  {\mathcal L}_{\rm CMB}(\eta,Y_p) \
  {\mathcal L}_{\rm BBN}(\eta;Y_p) \ dY_p \, ,
  \label{CMB-BBN-eta}
\eeq
Finally, we can include the light-element abundances in the convolution, 
\beq
{\mathcal L}_{\rm CMB-BBN-OBS}(\eta) \propto \int 
{\mathcal L}_{\rm CMB}(\eta,Y_p)
  {\mathcal L}_{\rm BBN}(\eta;X_i) \
  {\mathcal L}_{\rm OBS}(X_i) \ \prod_i dX_i \, ,
  \label{CMB-OBS-eta}
\eeq
which gives the tightest constraint on $\eta$.

In a similar way, we can generalize the BBN and CMB likelihoods for the case when $N_\nu$ is allowed to vary   
leading to $\mathcal{L}_{\rm NCMB}$ and $\mathcal{L}_{\rm NBBN}$. These can be used in similar convolutions as the likelihood functions defined above. 
For further details, see \citet{Cyburt:2001pp,Cyburt:2015mya,Fields:2019pfx}.

\section{New LBT \he4 Observations}
\label{sec:new}

Improvements to the accuracy in the determination of the helium abundance used here stem directly from the dedicated LBT $Y_{\rm p}$ project. This is paper V in the series of papers presenting the results of this project. For an overview of the project, see paper I \citep{Skillman2025}. Papers II and III present our new LBT MODS spectra, and our new LBT NIR spectra, respectively \citep{Rogers2025,Weller2025}. The details of the analysis in determining $Y_{\rm p}$ are given in Paper IV \citep{Aver2025}.
Here we provide a brief summary of the project and the methods used to determine $Y$ for the \HII\ regions observed. 

The LBT $Y_{\rm p}$ project is based on a set of observations of 54 extragalactic \HII\ regions chosen with a cut on metallicity, with O/H $< 15 \times 10^{-5}$, and sufficiently high signal-to-noise spectra. For each system, we establish a $\chi^2$ likelihood distribution summed over each observed emission line with wavelength $\lambda$, 
\beq
\chi^2 = \sum_{\lambda} \frac{\left(\frac{F(\lambda)}{F({\rm H}\beta|{\rm P}\gamma)} - {\frac{F(\lambda)}{F({\rm H}\beta|{\rm P}\gamma)}}_{meas}\right)^2} { \sigma(\lambda)^2},
\label{eq:X2}
\eeq
where the emission line fluxes, $F(\lambda)$, normalized to H$\beta$ (or the Paschen $\gamma$ line), are measured and calculated for a set of H and He lines, and $\sigma(\lambda)$ is the measured uncertainty in the flux ratio at each wavelength. Ideally, eight helium emission line fluxes are measured along with 15 hydrogen lines and one blended line, thus summing over 24 wavelengths in Eq.~(\ref{eq:X2}). Note that not every object had all 24 lines observed \citep[see][for details]{Aver2025}. 

The likelihood function (\ref{eq:X2}), is used to fit nine physical parameters: the electron temperature, $T_e$, electron density, $n_e$, optical depth, $\tau$, a reddening parameter, $C({\rm H}\beta)$, parameters for underlying absorption of He, and H Balmer/Paschen lines, $a_{\rm He}$, $a_{\rm H}$, and $a_{\rm P}$, a parameter for collisional excitation, $\xi$, and the ionized helium abundance (by number), $y^+$. Central values of the parameters are determined by minimization of the likelihood function. The uncertainties are determined by a Markov Chain Monte Carlo exploration of the parameter space. In the case of valid observations for all 24 emission lines, there are then 15 degrees of freedom. We impose a cut at the 95\% CL implying that only objects with $\chi^2 < 25.0$ are deemed reliable with respect to the model being used. Seven objects fail this test, leaving us with 47 \HII\ regions for further analysis. Note that in some of the objects which did not pass the 95\% cut, unusual features in the spectra are seen which may be particular to the individual \HII\ region, such as a nearby supernova, and we would not expect the model to be accurate in those cases. 
In addition, there are six objects which have been flagged due to an unusually high optical depth, underlying He absorption, or large discrepancies between the helium and oxygen temperatures. 
Thus we are left with a sample of 41 objects with ${\rm O/H} \le 15 \times 10^{-5}$. For more information on the model and cuts see Paper IV of this series \citep{Aver2025}.

The entire sample is at relatively low metallicity. 
Using the entire sample, there is very little evidence for a correlation between the helium abundance and metallicity (O/H).
This is to be expected for sufficiently low metallicity if the dispersion in the data exceeds $\Delta Y = Y - Y_{\rm p}$ at a given value of O/H. 
This is indeed the case for our sample, where a regression over the 41 points gives $Y = 0.2451 \pm 0.0022 + (14 \pm 12){\rm O/H}$. At sufficiently low metallicity, it is more appropriate to take the mean value of the lowest metallicity objects. In fact, the mean value is essentially unchanged within the uncertainties 
(as points with higher metallicity are added in) up to O/H $< 8 \times 10^{-5}$ (33 objects). On the basis of an F-test (for $\chi^2$ distributions), there is a 70\% chance that $Y$ is better determined by the mean than a regression (for which the slope is only determined to be $ 15 \pm 36$). The mean of these 33 points is $\langle Y \rangle = 0.2464 \pm 0.0013$. Compared to the BBN value of 0.2467, this shows remarkable agreement (in spite of the uncertainty). However, to be sure that we are not subject to the effects of evolution, we have chosen to determine the mean based on the lowest metallicity points with O/H $< 4 \times 10^{-5}$ (15 points) giving the result in Eq.~(\ref{Ypnew}). It is interesting to note that because of the dispersion in the data, despite using fewer points, the uncertainty in the mean is unchanged, but we gain in minimizing any (unaccounted for) effects of stellar contamination. 

\section{Results}
\label{sec:results}

\begin{figure*}[!htb]
\centering
\includegraphics[width=0.85\textwidth]{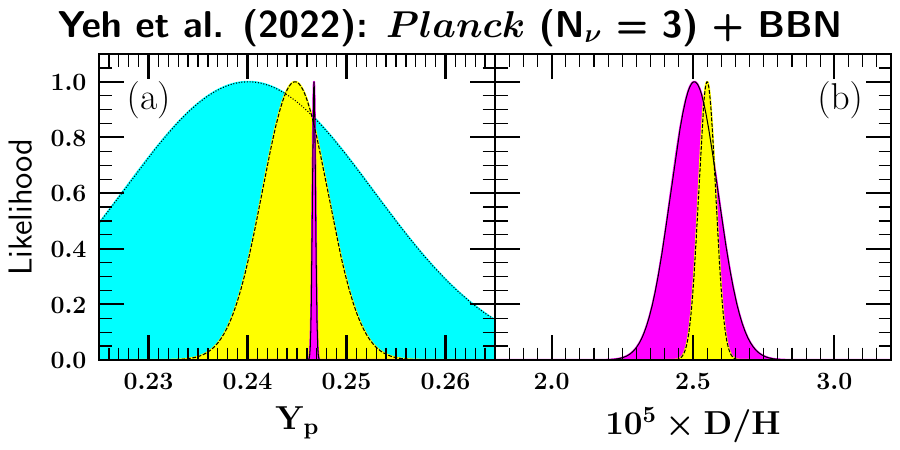}
\includegraphics[width=0.85\textwidth]{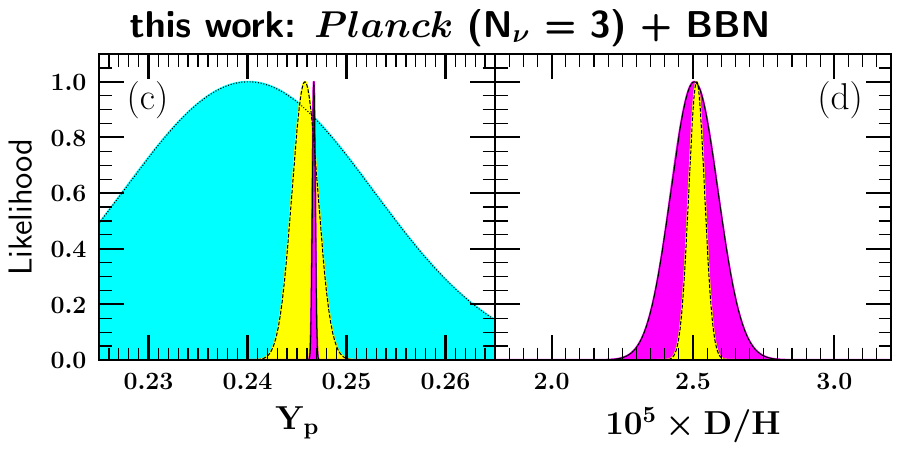}
\caption{The likelihood functions for $\Yp$ (left) and D/H (right).  The upper panels are taken from the analysis in \citet{Yeh:2022heq}, and are compared with those derived here, shown in the lower panels. 
BBN+CMB predictions are shown in the dark-shaded (purple) solid curves, and use {\em Planck} inputs.  Observational determinations for the $Y_{\rm p}$ and D/H primordial abundances appear as light-shaded (yellow) dashed curves. 
The independent CMB determination of \he4 appears as the medium-shaded (cyan) dotted curve.
\label{fig:2x2abs_2d}}
\end{figure*}

Using the formalism briefly outlined in Section \ref{sec:SBBN}, we have updated the BBN code using the new neutron mean life given in Eq.~(\ref{newtaun}) to establish new BBN likelihood functions. 
These are compared with observational likelihood functions for $Y_{\rm p}$ and D/H where the likelihood, $\mathcal{L}_{\rm obs}$ is simply a Gaussian with a central value and width given by either Eqs.~(\ref{DHnew}) or (\ref{Ypnew}). 
Figure~\ref{fig:2x2abs_2d} shows
a comparison of the likelihood functions derived in \citet{Yeh:2022heq} with $\Yp$ taken from \citet{Aver:2021rwi} (upper panels) with the likelihood functions derived here (lower panels) for (a,c) $Y_p$ (left) and (b,d) D/H (right). In the case of \he4, we also show the likelihood function defined in Eq.~(\ref{CMByp}). The combined CMB-BBN likelihood from Eq. (\ref{CMB-BBN}), ${\mathcal L}_{\rm CMB-BBN}(Y)$, is shaded purple. The observational likelihood 
is shaded yellow. The CMB-only likelihood, is shaded cyan.

The current CMB-BBN likelihoods (purple) in lower panels of Fig.~\ref{fig:2x2abs_2d}
are summarized by
the predicted abundances\footnote{Note that for \he4, our result including an additional significant digit is $Y_p = 0.24670 \pm 0.00014$ with a peak value of 0.24671. However given the residual systematic uncertainties in the calculation, we are not confident in the 5th decimal place and we have rounded up the uncertainty to 0.0002.}
\beqar
Y_p &=& 0.2467\pm0.0002 \qquad \qquad (0.2467) \label{meyp} \\
{\rm D/H} &=& (2.506\pm0.083)\times 10^{-5} \qquad (2.504 \times 10^{-5})\label{medh}\\
\he{3}/{\rm H} &=& (10.45\pm 0.87)\times 10^{-6} \qquad (10.45 \times 10^{-6}) \label{mehh} \\
\li{7}/{\rm H} &=& (4.95\pm 0.70)\times 10^{-10} \qquad (4.95 \times 10^{-10}) \label{melh}
\eeqar
where the central values give the mean,
and the error gives the $1\sigma$ variance.
The final number in parentheses gives the value at the peak of the distribution.

As one can plainly see, compared with the previous analysis shown in the upper panels, the biggest change in the \he4 comparison is the dramatic thinning of the observational
likelihood function (from an uncertainty of 0.0033 to 0.0013)
and some movement towards the CMB-BBN peak likelihood value of 0.2467 (from 0.2449 to 0.2458). Ideally, we would like to make the observational uncertainty more competitive with the theoretical one. The Planck CMB-only determined value is $Y_{\rm p} = 0.239 \pm 0.013$ and now lags far behind the direct observational uncertainty. We note that some progress has been achieved by ground-based CMB measurements, notably by Atacama Cosmology Telescope (ACT) \citep{ACT:2025tim} which finds 
 $Y_{\rm p}$ $= 0.2312 \pm 0.0092$ and the South Pole Telescope  has obtained $Y_{\rm p}$ $= 0.2285 \pm 0.0085$ (68\% CL) when its data is combined with Planck and ACT data \citep{SPT-3G:2025bzu}.

 The comparison of the D/H likelihood functions is almost perfect as can be expected by comparing Eqs.~(\ref{DHnew}) and (\ref{medh}). In this case, improvements to the theoretical cross sections used in BBN calculations are needed to make the predicted accuracy competitive with the observations from quasar absorption systems. As explained earlier, we make no direct comparison of the \he3 and \li7 likelihood functions with observation.

To appreciate the effect of the new value of $Y_{\rm p}$ on the determination of the baryon density, we provide in Table~\ref{tab:eta} the mean (with its uncertainty) and peak value of the determination of $\eta$ from the CMB alone, BBN (with and without the CMB), and using either or both $Y_{\rm p}$ and D/H.  Our preferred result is the last line of the table which combines the CMB, BBN, and observational likelihoods. We also use Eq.~(\ref{ombeta})
to provide the values of $\Omega_{\rm B} h^2$. The slight dependence on $Y_p$ does not affect the accuracy displayed. Note that these values are hardly changed from those given in \cite{Fields:2019pfx}, as \he4 plays only a minor role in determining $\eta$. This is evidenced in the 2nd row labeled BBN$+Y_p$ which shows a significantly larger uncertainty when either the CMB or D/H are not used in the analysis.  The slight changes seen here stem primarily from the updated value of D/H used in Eq.~(\ref{DHnew}).

\begin{table*}[!htb]
\caption{ Determination of the baryon-to-photon ratio, using different combinations of observational constraints.  We have marginalized over $Y_p$ to create 1D $\eta$ likelihood distributions with $N_\nu = 3$. The mean value (and its uncertainty) is given along with the value of $\eta$ at the peak of the distribution. In the final two columns, we have used Eq.~(\ref{ombeta}) to determine $\Omega_{\rm B} h^2$. The slight dependence on $Y_p$ does not affect the number to the displayed accuracy for reasonable values of $Y_p$. 
\label{tab:eta}
}
\vskip .2in
\begin{center}
\begin{tabular}{|l||c|c||c|c|}
\hline
 Constraints Used & mean $10^{10} \eta$ & peak $10^{10} \eta$ & mean $\omb$ & peak $\omb$ \\
\hline
CMB-only & $6.104\pm 0.055$ & 6.104 & $0.02230 \pm 0.00020$ & 0.02230 \\
\hline
\hline
BBN+$Y_p$ &$5.744^{+0.636}_{-0.875}$ & $5.563$ & $0.02098^{+0.00232}_{-0.00320}$& 0.02032 \\
\hline
BBN+D & $6.096\pm 0.119$ & 6.096 & $0.02227 \pm 0.00043$ & 0.02227 \\
\hline
BBN+$Y_p$+D & $6.089\pm 0.118$ & 6.088 & $0.02224 \pm 0.00043 $ & 0.02224 \\
\hline
CMB+BBN & $6.124\pm 0.040$ & 6.124  & $0.02237 \pm 0.00015$ & 0.02237 \\
\hline
CMB+BBN+$Y_p$ & $6.123\pm 0.040$ & 6.123 & $0.02237 \pm 0.00015$ & 0.02237 \\
\hline
CMB+BBN+D & $6.121\pm 0.038$ & 6.121 & $0.02236 \pm 0.00014$ & 0.02236 \\
\hline
\hline
CMB+BBN+$Y_p$+D & $6.120\pm 0.038$ & 6.120 & $0.02236 \pm 0.00014$ & 0.02236 \\
\hline
\end{tabular}
\end{center}
\end{table*}

\begin{figure*}[!htb]
\centering
\includegraphics[width=0.85\textwidth]{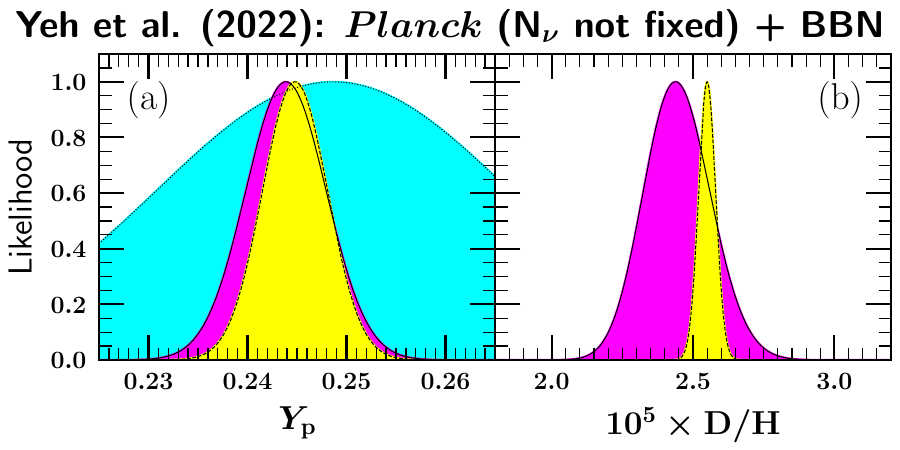}\\
\includegraphics[width=0.85\textwidth]{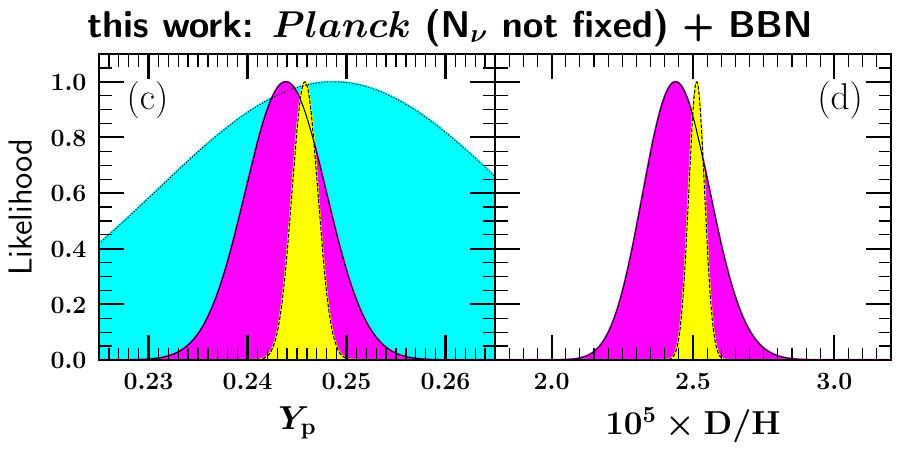}
\caption{ As in Fig.~\ref{fig:2x2abs_2d}, the
light element abundance likelihood functions using the CMB determination of the 
cosmic baryon density when $N_\nu$ is not fixed.   
}
\label{fig:2x2abs_3d}
\end{figure*}

 We can also perform the same analysis without fixing $N_\nu = 3$. In this case, $N_\nu$ is treated as a free parameter in both the CMB and BBN likelihood functions.
 For example, the CMB likelihood becomes 
 \beq
{\mathcal L}_{\rm NCMB}(\eta,N_\nu) \propto \int 
  {\mathcal L}_{\rm CMB}(\eta,N_\nu,Y_{\rm p}) \ dY_{\rm p} \, ,
  \label{LNCMB}
\eeq
with a similar expression for the BBN likelihood function when $N_\nu$ is allowed to vary. 
The convolved likelihood function is now given by 
\beq
{\mathcal L}_{\rm NCMB-NBBN}(X_i) \propto \int 
  {\mathcal L}_{\rm NCMB}(\eta,N_\nu,Y_p) \
  {\mathcal L}_{\rm BBN}(\eta,N_\nu;X_i) \ d\eta \ d N_\nu\, .
\label{NCMB-NBBN}
\eeq
 The comparison of this likelihood function (purple) with ${\mathcal L}_{\rm OBS} (X_i)$ (yellow) and ${\mathcal L}_{\rm NCMB}(Y_p)$ for \he4 (cyan) is shown in Fig.~\ref{fig:2x2abs_3d}. As in  Fig.~\ref{fig:2x2abs_2d}, we show the likelihood functions derived in \citet{Yeh:2022heq} in the upper panels and those derived in this work in the lower panels. The most notable change seen in this figure relative to Fig.~\ref{fig:2x2abs_2d} with $N_\nu = 3$ is the broadening of the NCMB-NBBN likelihood function (purple) for \he4 (and a slight shift to lower values). Indeed, when $N_\nu$ is {\em not} fixed, the uncertainty in the new LBT $Y_p$ determination of the \he4 abundance is now {\em smaller} than the theoretical uncertainty. Of course the two likelihoods remain in excellent agreement. The same trend is seen for D/H where the NCMB-NBBN likelihood also shifts to slightly lower values. The NCMB-NBBN likelihoods in Fig.~\ref{fig:2x2abs_3d}
are summarized by
the predicted abundances
\beqar
Y_p &=& 0.24387 \pm0.00408 \label{ypnnu} \qquad \qquad (0.24421) \\
{\rm D/H} &=& (2.447\pm0.117)\times 10^{-5} \qquad (2.438 \times 10^{-5}) \\
\he{3}/{\rm H} &=& (10.37\pm0.88)\times 10^{-6} \qquad (10.37 \times 10^{-6}) \\
\li{7}/{\rm H} &=& (5.03\pm 0.72)\times 10^{-10} \qquad (5.02 \times 10^{-10}) \, .
\eeqar

It is also possible to formulate full 2D likelihood functions of $N_\nu$ and $\eta$. In addition to ${\mathcal L}_{\rm NCMB}$ defined in Eq.~(\ref{LNCMB}), we can also define 
\begin{equation}
{\cal L}_{\rm NBBN+obs}(\eta, N_\nu)  \propto \int  {\cal L}_{\rm NBBN}(\vec{X} ; \eta,N_\nu) \ \prod_i {\cal L}_{{\rm obs}}(X_i)  \ dX_i \, , \label{eq:LBBN} \\ 
\end{equation}
\begin{equation}
\begin{array}{l}
   {\cal L}_{\rm NBBN+NCMB+obs}(\eta, N_\nu) 
    \propto  \\ 
    \int {\cal L}_{\rm NCMB}(\eta,N_\nu,Y_p) \ 
    {\cal L}_{\rm NBBN}(\vec{X};\eta, N_\nu) \ 
    \prod  {\cal L}_{\rm obs}(X_i) \ dX_i  \ \ .  
    \end{array}
    \label{BBNCMBobs}
\end{equation}
From these, we can obtain 1D likelihood functions of either $\eta$ or $N_\nu$ by marginalizing over $N_\nu$ or $\eta$ respectively. 

The mean and peak values of $\eta$ and $N_\nu$ are summarized in Table~\ref{tab:etannu}. In the first four columns of values, we have marginalized the various likelihood functions in Eqs.~(\ref{LNCMB}), (\ref{eq:LBBN}), and (\ref{BBNCMBobs}) over $N_\nu$ to obtain the mean and peak values of $\eta_{10}$ and as before we used Eq.~(\ref{ombeta}) to determine $\Omega_{\rm B} h^2$. Overall, there is a slight downward shift in $\eta$ when $N_\nu$ is allowed to vary.
However the impact of the new observational determination of $Y_p$ increases slightly the best fit to $\eta$ and reduces its uncertainty. In \citet{Yeh:2022heq}, it was found that on combining the CMB with BBN and the observations of $Y_p$ and D/H, $\eta_{10} = 6.088 \pm 0.054$ using $Y_p = 0.2448 \pm 0.0033$ from \citet{Aver:2021rwi}. This is to be compared with the last line in Table \ref{tab:etannu} based on $Y_p$ given by Eq.~(\ref{Ypnew}) and yields an improved accuracy in the determination of $\eta_{10}$ by roughly 20\%.

\begin{table*}[!htb]
\caption{{\bf to be checked} The separately marginalized central 68.3\% confidence limits and most-likely values on the baryon-to-photon ratio $\eta$ and effective number of neutrinos $N_\nu$, using different combinations of observational constraints. The 95.45\% upper limits from Eq.~(\ref{int}), given that $ N_{\nu}>3$, are also shown in the last column.}
\label{tab:etannu}
\vskip.1in
\begin{center}
\begin{tabular}{|l||c|c||c|c||c|c||c|}
\hline
 Constraints Used & mean $\eta_{10}$ & peak $\eta_{10}$ & mean $\omb$ & peak $\omb$ &mean $N_\nu$ & peak $N_\nu$ & $\delta N_{\nu}$ \\
\hline\hline
CMB-only & $6.090 \pm 0.061$ &  $6.090$ & $0.02225 \pm 0.00022$ & 0.02225 & $2.800 \pm 0.294$ &  $2.764$ & 0.513\\
\hline                  
BBN+$Y_p$+D & $6.066 \pm 0.122$ &  $6.065$ & $0.02216 \pm 0.00045 $& 0.02215 & $2.941 \pm 0.092$ &  $2.939$ & 0.151\\
\hline                   
CMB+BBN & $6.087 \pm 0.061$ &  $6.088$ & $0.02224 \pm 0.00022$ & 0.02224 & $2.848 \pm 0.190$ &  $2.843$ & 0.298\\
\hline                  
CMB+BBN+$Y_p$ & $6.105 \pm 0.044$ &  $6.105$ & $0.02230 \pm 0.00016$ & 0.02230 & $2.917 \pm 0.084$ &  $2.916$ & 0.125\\
\hline                   
CMB+BBN+D & $6.090 \pm 0.060$ &  $6.091$ & $0.02225 \pm 0.00022$ & 0.02225 & $2.886 \pm 0.176$ &  $2.882$ & 0.288\\
\hline                    
CMB+BBN+$Y_p$+D & $6.101 \pm 0.044$ &  $6.102$ & $0.02229 \pm 0.00016$ & 0.02229 & $2.925 \pm 0.082$ &  $2.924$ & 0.125\\
\hline                    
\end{tabular}
\end{center}
\end{table*}

\begin{figure*}[!ht]
    \centering
    \includegraphics[width=\textwidth]{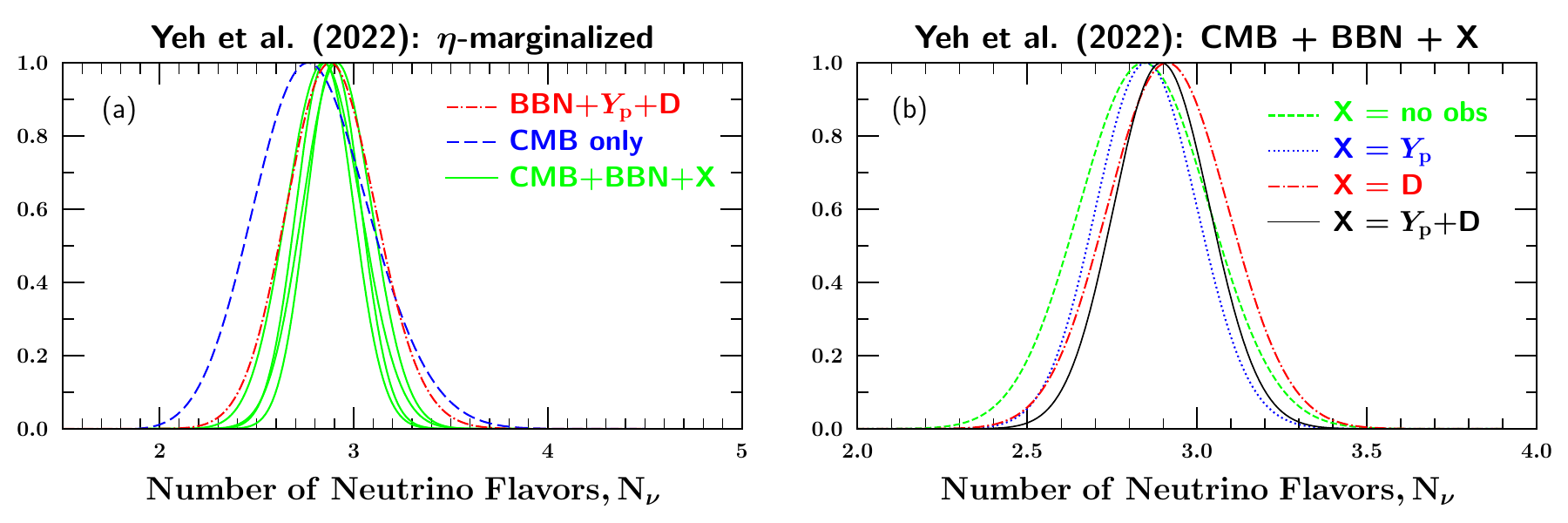}
    \includegraphics[width=\textwidth]{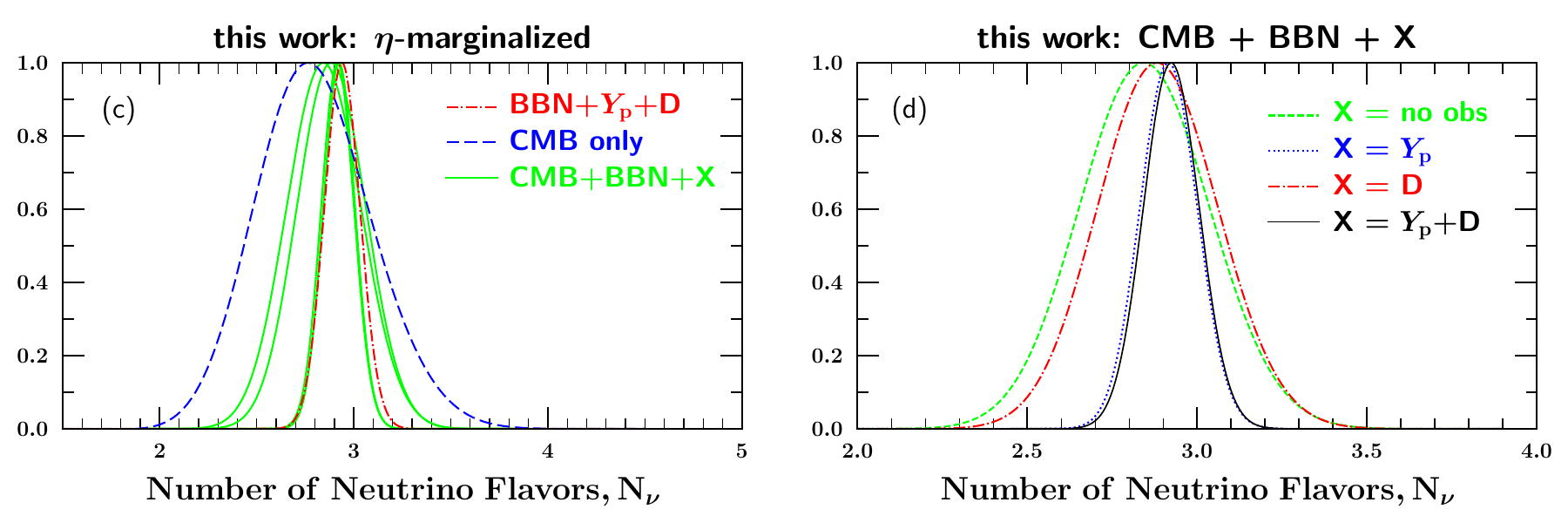}
    \caption{ Likelihood distributions for $N_\nu$ using various combinations of 
    BBN and CMB data.
    The initial likelihood functions have been marginalized over the baryon-to-photon ratio $\eta$.
    Older results based on $Y_p = 0.2448 \pm 0.0033$ from \citet{Aver:2021rwi} are shown in panels (a,b) while results from this work using Eq.~(\ref{Ypnew}) are shown in panels (c,d). 
    We show in panels (a,c) the likelihood distributions for BBN-only, CMB-only, and combined limits. In panels (b,d) we zoom into BBN+CMB joint limits to highlight results for different combinations of light element abundances. The last three columns of
    Table \ref{tab:etannu} summarize these results.}
    \label{fig:N_nu_dist_Aver}
\end{figure*}

By marginalizing the likelihood functions (\ref{LNCMB}), (\ref{eq:LBBN}), and (\ref{BBNCMBobs}) over the baryon-to-photon ratio, we can obtain the various likelihood function for $N_\nu$. 
To see more clearly the impact of the new observational determination of $Y_p$, we compare our current result with previous results from \citet{Yeh:2022heq}. 
These are shown in Fig.~\ref{fig:N_nu_dist_Aver}.
The upper panels show the result based on $Y_p = 0.2448 \pm 0.0033$ from \citet{Aver:2021rwi} while the lower panels show our current results. 
The left panels of this figure show $\mathcal{L}_{\rm NCMB}(N_\nu)$ as the blue dashed curve using Planck data alone. 
The red dot-dashed curve shows the $\mathcal{L}_{\rm NBBN-OBS}$ likelihood using both $Y_{\rm p}$ and D/H and the series of solid green curves show the $\mathcal{L}_{\rm NCMB-NBBN-OBS}$ with various combinations of the light elements included. For clarity, these are displayed (and labeled) separately in the right panel. 
Note that the case labeled `$X =$ no obs' 
shows the likelihood $\int {\cal L}_{\rm NCMB-NBBN}(Y_p; \eta,N_\nu) \ d\eta \ dY_p$
where no observational data are included.
Even without the element observations, we obtain a stronger
limit on $N_\nu$ than
possible using the CMB alone.  
In comparing the old (upper panels) and new (lower panels) results, we see a remarkable improvement in the width of the likelihood distribution when \he4 is included.

The mean and peak values of $N_\nu$ for all of the likelihood functions shown lie slightly below the Standard Model value of $N_\nu = 3$. The precise values are listed in Table~\ref{tab:etannu}. However, they are all within 1$\sigma$ of the Standard Model value. Clearly, the 
BBN and CMB determinations of $N_\nu$ are in excellent 
agreement with each other, and with the Standard Model value.
This conclusion makes use of the implicit assumption that $N_\nu^{\rm BBN} = N_\nu^{CMB}$.
New physics beyond the Standard Model could cause these two measures of the number of relativistic degrees of freedom to differ, and their agreement points further to the lack of new physics affecting this number. 

From Fig.~\ref{fig:N_nu_dist_Aver} and Table~ \ref{tab:etannu}
we see that $Y_p$ has a strong impact on the constraints on $N_\nu$. 
Our best limit (last line in Table~\ref{tab:etannu}) uses both light elements and the CMB:
\begin{equation}
\label{eq:Nnu-tout}
    N_\nu = 2.925 \pm 0.082 \, .
\end{equation}
This gives a 2$\sigma$ upper limit of
\begin{equation}
\label{eq:DeltaNnu2side}
    \Delta N_\nu = N_\nu - 3 < 0.089
\end{equation}
arising from our two-sided error range about the mean in Eq.~(\ref{eq:Nnu-tout}).
This result updates that given in \cite{Yeh:2022heq} ($N_\nu = 2.898 \pm 0.141$ or $\Delta N_\nu < 0.180$)
by the inclusion of the updated D/H abundance,  updated neutron mean-life, and most importantly, the new primordial helium abundance given in Eq.~(\ref{Ypnew}).
The first two of these updates are incremental. 
This upper limit is about a factor of 20 tighter than in \cite{Cyburt2005}.

Given that the best fit value of $N_\nu$ is slightly below 3, the resulting limit 2-$\sigma$ upper limit in Eq.~(\ref{eq:DeltaNnu2side}) may be overly aggressive in the Standard Model which includes three neutrinos in the radiation bath.
If we assume a prior, $N_\nu \ge 3$ we can renormalize the likelihood function \citep{Olive:1995fi} to find the 1-sided limit given by 
\begin{equation}
\frac{\int^{3 + \delta N_{\nu}}_3 {\cal L}(N_\nu) \ dN_\nu}{\int^{\infty}_3 {\cal L}(N_\nu) \ dN_\nu} = 0.9545
\label{int}
\end{equation}
The last column of Table \ref{tab:etannu} gives 95.45\% CL values for 
$\delta N_\nu$, of which the strongest is
\begin{equation}
\label{eq:DeltaNnu3}
    \delta N_{\nu} = N_{\nu} - 3 \ < \ 0.125 
\end{equation}
based on the combination of CMB+BBN+$Y_p$+D/H.
This is to be compared with the previous limit of $\delta N_\nu < 0.226$ \citep{Yeh:2022heq}.

The impact of this result is wide ranging in its ability to constrain physics beyond the Standard Model. 
For example, a scalar particle in equilibrium contributes $\Delta N_\nu = 4/7 = 0.57$,
so it existence is ruled out, unless it decouples well before neutrino decoupling \citep{Steigman:1979xp,Olive:1980wz}. Another common application of this limit pertains to possible right-handed interactions. Assuming the existence of three, light neutrino like states, $\nu_R$ which interact only through right-handed interactions mediated by a heavy gauge boson, $Z^\prime$, the limit in Eq.~(\ref{eq:DeltaNnu3}) would require that the ratio of the temperature of these neutrinos, relative to Standard Model neutrinos is $T_{\nu_R}/T_{\nu_L} < 0.45$. While this appears to be only a modest improvement in the limit from 0.52 \citep{Yeh:2022heq},
it has profound consequences. To achieve $T_{\nu_R}/T_{\nu_L} < 0.45$, these neutrinos must decouple at a temperature $T_{d\nu_R} > m_t$. That is, the energy density of the right-handed interactions must be diluted by the full spectrum of standard Model particles and even then we would obtain, $T_{\nu_R}/T_{\nu_L} \simeq 0.465$. This would place a minimal constraint on the scale of right-handed interactions of $M_{Z^\prime} \gtrsim 350$~TeV!  For further details and other applications see \citet{Yeh:2022heq}. 

We do note that some of the strength in this result is tied to the fact that our central value for $N_\nu < 3$. For example, had we used $\Yp = 0.2464 \pm 0.0013$ (based on the mean of the 33 points with O/H $< 8 \times 10^{-5}$), we would have obtained $N_\nu = 2.959 \pm 0.082$ and $\delta N_\nu < 0.142$, thus slightly larger than the value given in Eq.~(\ref{eq:DeltaNnu3}). But even in this case, we would require $T_{\nu_R}/T_{\nu_L} < 0.466$ and would still require
full dilution from all Standard Model degrees of freedom. 

\begin{figure*}
\centering  \includegraphics[width=\textwidth]{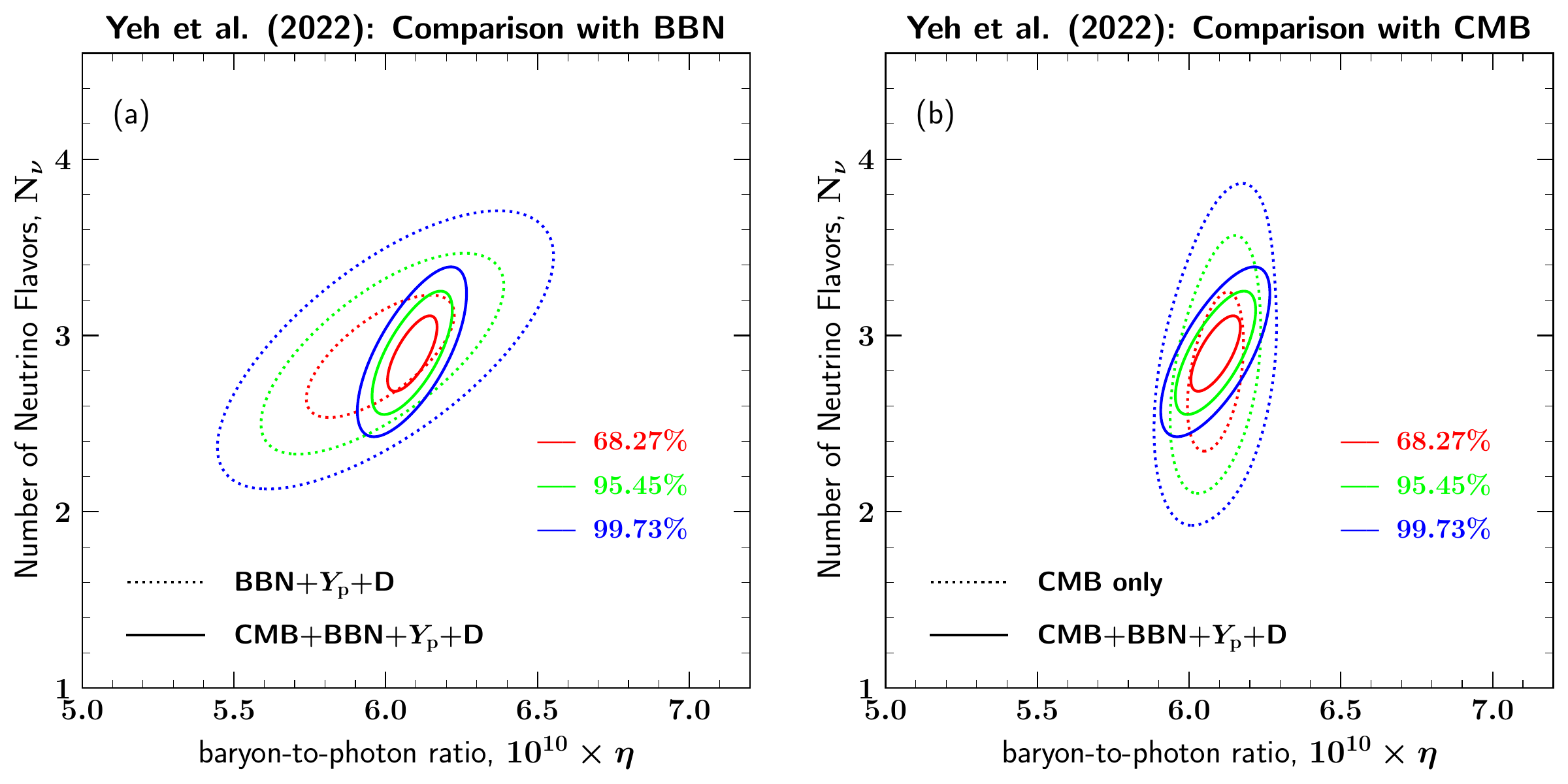}
    \centering  \includegraphics[width=\textwidth]{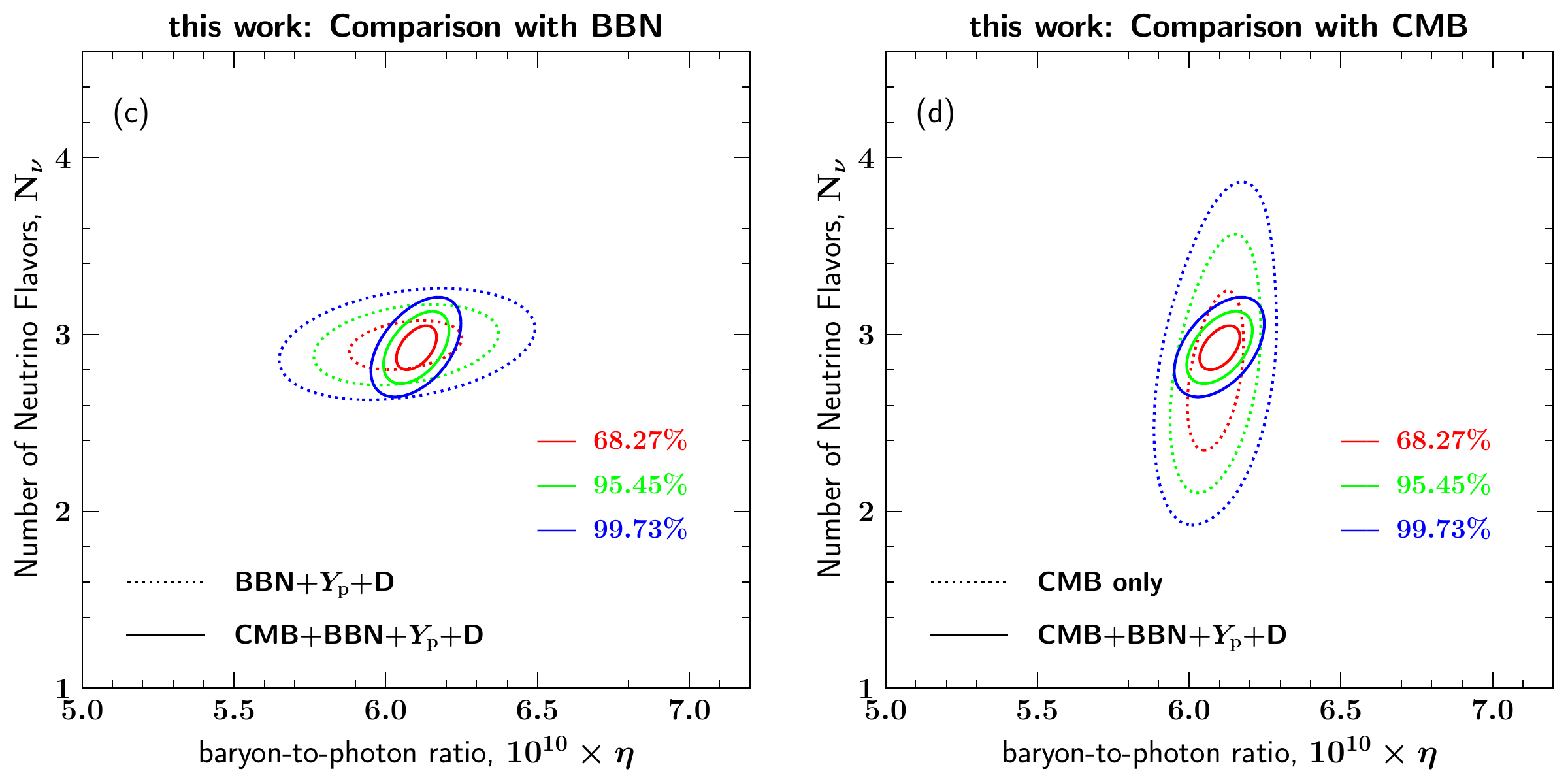}
    \caption{The 2D likelihood ${\cal L}(\eta,N_\nu)$.  
    Older results based on $Y_p = 0.2448 \pm 0.0033$ from \citet{Aver:2021rwi} are shown in panels (a,b) while results from this work using Eq.~(\ref{Ypnew}) are shown in panels (c,d). 
   In panels (a,c) the dotted contours show BBN-only results, while
    solid contours give combined BBN+CMB results. 
    In panels (b,d) the dotted contours correspond to the CMB-only likelihood, with solid contours showing again the BBN+CMB combined result. 
    }
    \label{fig:Nnu_vs_eta}
\end{figure*}

The full 2D likelihood functions can be visualized as 
a projection onto the $(\eta, N_\nu)$ plane.  
These are shown in Fig.~\ref{fig:Nnu_vs_eta}, where again we compare the previous results in \citet{Yeh:2022heq} using $Y_p = 0.2448 \pm 0.0033$ in the upper panels with the results from this work using $Y_P$ from Eq.~(\ref{Ypnew}). The combined NCMB-NBBN-OBS likelihood function in both cases 
is shown by the solid contours. 
The shrinking of these ellipses is almost entirely due the improvement in the uncertainty in $Y_p$. 
These likelihoods are compared with the NBBN-only results using Eq.~(\ref{eq:LBBN}) and the NCMB-only results using Eq.~(\ref{LNCMB}) in the left and right panels shown by the dotted contours. 
The striking agreement seen by the overlapping contours of the BBN-only, CMB-only, and combined results indicates 
the concordance between these two critical tests of cosmology. 
As expected, and clearly seen in
Figure~\ref{fig:Nnu_vs_eta}, BBN provides an accurate determination of $N_\nu$, whereas the CMB provides an accurate determination of the baryon density. 
Most importantly, these results are all in excellent agreement with the Standard Model value of $N_\nu=3$.

\section{Summary}
\label{sec:sum}

The early evolution of the universe is well described by Standard Big Bang Cosmology with cold dark matter and a cosmological constant ($\Lambda$CDM) and whose fundamental parameters are extremely well determined by measurements of the anisotropies in the CMB \citep{Planck:2018vyg}. This establishes the state of the Universe at a red-shift of $z\sim 1100$. To push our knowledge of the early universe back to $z \sim 10^{10}$ or energy scales of $\sim 1$~MeV so that Standard Model physics can be tested, we rely on the success (or not) or big bang nucleosynthesis. However, the precision testing of cosmology at this epoch is dependent on 1) accurate cross-section measurements involving the light element isotopes, and 2) accurate abundance determinations. 

As clearly seen by the purple shaded likelihood functions in Fig.~\ref{fig:2x2abs_2d}, current experimental cross-sections lead to very accurate predictions of \he4, with an uncertainty in $\Yp$ of less than 0.0002 (i.e., $< 0.1\% $), and a reasonably well predicted D/H abundance with an uncertainty of $0.083 \times 10^{-5}$ (i.e., $ \lesssim 3\%$ ). In the case of deuterium, abundance determinations from quasar absorption systems are at the $\sim 1\% $ level, a factor of 3 better than theory. On the theory side there is clearly room for improvement with new measurements of the d(d,p) and d(d,n) cross-sections \citep{2021Yeh}. In the case of \he4, the observational uncertainties have historically lagged behind the theoretical uncertainties. In \citet{aver2015}, it was found that by adding the infra-red line $\lambda$10830, the uncertainty in $\Yp$ could be as low as 0.0040, a 1.6\% uncertainty, which while competitive with D/H is far greater than the theoretical uncertainty. 

In this series of papers \citep{Skillman2025,Rogers2025,Weller2025,Aver2025} including the present work, we concentrated on obtaining
many high quality observations at low metallicity (O/H $< 4 \times 10^{-5}$). The result of this work led to the determination of the primordial \he4 abundance given in Eq.~(\ref{Ypnew}) with an uncertainty at the $\sim 0.5 \%$ level. 
It is based on the observation of only 15 objects and lends promise to a further reduction in the uncertainty with more such high-quality observations. 

The accurate determinations of \he4 and deuterium allow us to test standard big bang cosmology at an age of $\sim 1$~sec
at an unprecedented level. At the CMB-determined value of the baryon density, we see clearly in Fig.~\ref{fig:2x2abs_2d}
the agreement between the BBN predictions and astronomical observations. 
 
This concordance holds when we allow the number of neutrino flavors (or other particle degrees of freedom) to vary from $N_\nu = 3$. 
Note that in this case, as seen in Fig.~\ref{fig:2x2abs_3d}, our $\Yp$ determination is now more accurate than the combined CMB+BBN prediction. The latter uncertainty is $\sigma(Y_{\rm CMB+BBN}) = 0.00408$, or only a 1.7 \% prediction.

We have shown here that by combining the CMB, BBN, and observational likelihoods we are able to improve on the CMB determination of the baryon density by reducing the uncertainty in $\eta_{10}$ from .061 to .044. Perhaps more importantly, the combination of the likelihoods leads to a mean value for $N_\nu = 2.925 \pm 0.082$ to be compared with the Standard Model value of 3. The small uncertainty, resulting primarily from the reduced uncertainty in $\Yp$,
places strong constraints on physics beyond the Standard Model. These constraints have implicitly assumed that the BBN and CMB values of $\eta$ and $N_\nu$ are the same. 
The concordance we have found shows no indication that this assumption is false. To the contrary, one can show \citep[as we did in][]{Yeh:2022heq}, there are strong constraints on any possible variation in $\eta$ and $N_\nu$ between these two epochs. That conclusion is maintained with the improved value of $\Yp$. 

\begin{figure*}
    \centering
\includegraphics[width=0.85\textwidth]{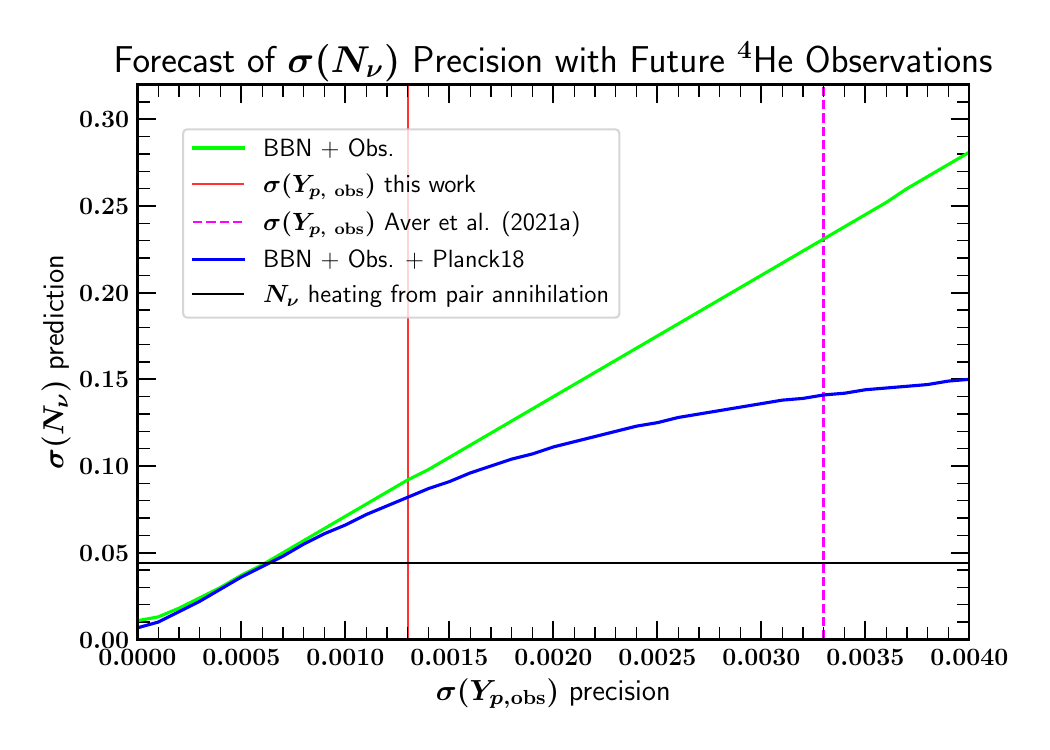}
    \caption{
    Forecast of expected precision in the measurement of $N_\nu$, shown as a function of the precision of the $Y_{\rm p,obs}$ measurement. The green curve shows the precision of BBN-only measurements, which show a nearly linear scaling as in eq.~(\ref{eq:sig-scaling}).
    The blue curve shows the improvement when adding {\em Planck} CMB measurements.  Vertical lines are the previous (magenta) and present (red) $Y_{\rm p,obs}$ precision.  We see that an additional factor $\sim 2$ improvement in $\sigma(Y_{\rm p,obs})$ promises to bring the $N_\nu$ precision around the value due to neutrino heating in the early Universe, shown in the horizontal line.
    }
    \label{fig:futureYp}
\end{figure*}

As we noted earlier, for $N_\nu = 3$, the effective number of degrees of freedom contributing to the expansion rate post BBN is $N_{\rm eff} = 3.044$ \cite{Drewes:2024wbw}. 
Our ultimate goal, therefore would be to bring down the uncertainty in $\Yp$ so that the difference between $N_\nu$ and $N_{\rm eff}$ can be tested. We show in Fig.~\ref{fig:futureYp}, the necessary accuracy in $\Yp$, $\sigma(\Yp)$ needed to obtain a given accuracy in $N_\nu$, $\sigma(N_\nu)$. 
The BBN-only dependence of $\sigma(N_\nu)$ on $\sigma(\Yp)$ can be determined from Eq.~(\ref{yfit}). This gives
\beq
\label{eq:sig-scaling}
\sigma(N_\nu) = (1/0.163)(N_\nu/\Yp) \, \sigma(\Yp) \approx 70 \, \sigma(Y_p) \, . 
\eeq
This linear relation is shown by the green line in  Fig.~\ref{fig:futureYp} and reflects
the fact that for BBN-only, $N_\nu$ is primarily determined by $\Yp$. 
To test the effects  of neutrino heating
with $\sigma(N_\nu) < 0.044$ (shown by the solid black line) requires a
very precise helium determination:
$\sigma(Y_p) < 0.0006$. But this is only a factor of 2 better than the current uncertainty!
The current value of $\sigma(\Yp)$ is depicted by the vertical red solid line at $\sigma(\Yp) = 0.0013$. This should be compared with the previous value at $\sigma(\Yp) = 0.0033$ \citep{Aver:2021rwi} shown by the vertical dashed blue line. 
Future improvements in CMB measurements can provide competitive measurements:  
the Simons Observatory is projected to measure $N_{\rm eff}$ to a precision reaching $\sigma(N_{\rm eff}) = 0.045$ \citep{Abitbol2025SimonsObs}.  This would allow a spectacularly precise test of cosmology and early universe physics.

In the past, it was seen that adding in the Planck CMB measurements makes a dramatic improvement in $\sigma(\Yp)$.
For example at the position of the vertical blue dashed line, including the Planck likelihood, reduces $\sigma(N_\nu)$ from 0.23 to 0.14 as seen by comparing the green and blue curves in Fig.~\ref{fig:futureYp}. While there is still improvement when $\sigma(\Yp) = 0.0013$, the improvement is minimal, taking $\sigma(N_\nu)$ from $\sim 0.09$ to $\sim 0.08$. 
Fully testing $\sigma(N_\nu) \le 0.044$ will require either 
a helium measurement at the (1/4)\% level, as mentioned above or a new CMB analysis sensitive enough to probe $N_\nu$
at that accuracy.

\begin{acknowledgements}
BDF is pleased to acknowledge useful discussions with Chinmay Ambasht, Gil Holder, and Cynthia Trendafilova.
TRIUMF receives federal funding via a contribution agreement with the National Research Council of Canada. The work of KAO is supported in part by DOE grant DE-SC0011842 at the University of
Minnesota. This work was supported by funds provided by NSF Collaborative Research Grants AST-2205817 to RWP, AST-2205864 to EDS, and AST-2205958 to EA. 

This work is based on observations made with the Large Binocular 
Telescope. The LBT is an international collaboration among institutions 
in the United States, Italy and Germany. LBT Corporation Members are: 
The University of Arizona on behalf of the Arizona Board of Regents; 
Istituto Nazionale di Astrofisica, Italy; LBT Beteiligungsgesellschaft, 
Germany, representing the Max-Planck Society, The Leibniz Institute for 
Astrophysics Potsdam, and Heidelberg University; The Ohio State 
University, and The Research Corporation, on behalf of The University 
of Notre Dame, University of Minnesota and University of Virginia. 
Observations have benefited from the use of ALTA Center 
(alta.arcetri.inaf.it) forecasts performed with the Astro-Meso-Nh 
model. Initialization data of the ALTA automatic forecast system come 
from the General Circulation Model (HRES) of the European Centre for 
Medium Range Weather Forecasts.

This research used the facilities of the Italian Center for Astronomical Archive (IA2) operated by INAF at the Astronomical Observatory of Trieste.

EDS, KAO, EA, and NSJR would like to acknowledge and thank Stanley 
Hubbard for his generous gift to the University of Minnesota that 
allowed the University to become a member of the LBT collaboration.

\end{acknowledgements}

\bibliographystyle{aasjournalv7}
\bibliography{He4-V}

\clearpage

\end{document}